\documentclass[letterpaper]{article}
\usepackage{aaai}
\usepackage{times}
\usepackage{helvet}
\usepackage{courier}
\usepackage{url}
\usepackage{multirow}
\usepackage[table,xcdraw]{xcolor}
\usepackage{float}
\usepackage{array,graphicx}
\usepackage{booktabs}
\usepackage{pifont}
\title{A Structured Analysis of Journalistic Evaluations for News Source Reliability}

\author{Manuel Pratelli\textsuperscript{\rm 1,2} \and Marinella Petrocchi\textsuperscript{\rm 1,2}\\
\textsuperscript{\rm 1} IMT Scuola Alti Studi Lucca\\
Piazza San Francesco 19, 55100, Lucca Italy\\
\and \\ \textsuperscript{\rm 2} IIT-CNR, via G. Moruzzi 1, 56124, Pisa, Italy\\
}

\begin{document}

%

\maketitle
\begin{abstract}
\begin{quote}
In today's era of information disorder, many news organizations are moving to verify the veracity of news published on the web and social media. In particular, some agencies are exploring the world of online media and, through a largely manual process, ranking the credibility and transparency of news sources around the world. In this paper, we evaluate two procedures for assessing the risk of online media exposing their readers to m/disinformation. The procedures have been dictated by NewsGuard and The Global Disinformation Index, two well-known organizations combating d/misinformation via practices of good journalism. Specifically, considering a fixed set of media outlets, we examine how many of them were rated equally by the two procedures, and which aspects led to disagreement in the assessment. The result of our analysis shows a good degree of agreement, which in our opinion has a double value: it fortifies the correctness of the procedures and lays the groundwork for their automation.
\end{quote}
\end{abstract}

\noindent 

\section{Introduction}
The term \textit{infodemic}, formed by the two words `information' and `epidemic', refers to `a rapid and far-reaching spread 
of both accurate and inaccurate information about something, 
such as a disease'\footnote{UN tackles `infodemic' of misinformation and cybercrime in COVID-19 crisis. March 31, 2020. Online: \url{https://www.un.org/en/un-coronavirus-communications-team/un-tackling-`infodemic'-misinformation-and-cybercrime-covid-19}. All URLs in this manuscript have been lastly accessed April 8, 2022.}. The term has been in the news since the end of March 2020: it was adopted by the United Nations and the World Health Organization to characterize the media and information disorder related to the Covid-19 pandemic. Clearly, between true and invented narratives, not to mention all shades of truth that there may be between one extreme and another, it becomes extremely difficult to discern quality and reliability of information. 

In order to help the readers to shed light on the nature of the magnum sea of online information to which they are exposed, many local, European and international initiatives have arisen. 
As an example, The European Digital Media Observatory (EDMO)\footnote{\url{edmo.eu}} brings together fact-checkers, academicians and communication experts to understand and identify disinformation, uproot its sources and dilute its impact. Thanks to the fact checkers of its network, EDMO regularly updates its website with indications of false news detected in the EU environment. This is, for instance, the page related to fact-checked disinformation on the war in Ukraine detected during 2022: 
\url{https://bit.ly/3pJVx7l}

There are also nonprofit organizations that combat d/misinformation through expert procedures of good journalism. NewsGuard (NG) -\url{newsguardtech.com}- is an organization of qualified journalists who evaluate news sources responsible for 95 percent of online engagement. In 2021, NewsGuard monitored thousands of news sites, revealing the publication of misleading information about vaccines and political elections, among others, and estimating \$2.6 billion in annual ad revenue going to sites responsible for disinformation. 

Similarly, The Global Disinformation Index (GDI) -\url{disinformationindex.org}- assesses a media site's overall risk of carrying disinformation, via a selection of transparency and credibility indicators and human reviews carried on by country-specific analysts. Among the reports published by the index, the interested reader can find a series of publications concerning the online media market for various countries, such as Canada, United States, Australia, many European countries, and some in Latin America and Africa, see, e.g.~\cite{mediamarketItaly2022} for a recent  analysis of the Italian news media, titled   `Disinformation Risk Assessment: The Online News Market in Italy'. 
The purpose of the two organizations is similar: to debunk misinformation, increase reader awareness, and attempt to divert advertising streams from low or no reputable news sites. Obviously, an analysis of this kind, whether it concerns newspapers or a specific topic (health, politics, internal and foreign affairs of a country...) requires a considerable human effort and, despite this painstaking work, it cannot achieve a complete media coverage. Evaluating all news sources is not physically feasible, hence the idea  of selecting those  characterized by the big amount of traffic.


Hence, the question that poses the bases for this manuscript:  \textit{Would it be worthwhile to automate the process?}
With an abuse of notation, what we mean is to build a sort of `automated fact checking on the news outlet', where what is evaluated is not the truthfulness of the single fact, or claim, or post --such as in, e.g., \cite{DBLP:journals/tweb/ZubiagaJ20,DBLP:journals/llc/ZengAZ21}-- but the transparency and credibility of the whole online news media. The automation of the process is particularly appealing for what concerns the evaluation of the `tail' of news sources (brand new ones or with less traffic than those considered by the actuators of the manual verification).

In order to lay a solid foundation for automating the process, in this paper we present an assessment of the  procedures and outcomes of the GDI and NewsGuard evaluation methodology, ran over the same set of news media.
An assessment of this kind  acquires considerable importance if we consider that, in the literature, researchers have conducted analysis in the field of misinformation by leveraging the tags of news sources assigned by such organizations, see, e.g.,~\cite{aker2019credibility,doi:10.1126/science.aau2706,Shao18,Mattei22bowties,CaldarelliNPPS21}. 
Despite the wide use of this approach, no one, to the best of our knowledge, has so far measured the agreement between different evaluation processes (agreement measured both in terms of criteria adopted and in terms of final scores given to the news sources).

Thus, the first research question we would like to answer here is: 

\textbf{RQ1 --} \textit{
Do the implementation of different decision processes lead to the same results in terms of estimating the reliability of online information sources?}

As introduced above, both organizations evaluate worldwide online media markets, based on a set of criteria inspired by good journalism principles and practices.
The considered criteria, while similar to each other, are obviously not the same. Thus, the procedures' assessment must necessarily address the issues that give rise to the second and third research questions of this paper:

\textbf{RQ2 --} \textit{Can we establish a conceptual mapping between the two sets of criteria?  In other words, is it possible to relate the criteria adopted by one organization to the criteria of the other, and vice versa?}

\textbf{RQ3 --} \textit{Focusing on the criteria that find conceptual mapping, and considering the result of their application on the same set of news sources, what are the criteria that obtain a concordant (resp., discordant) evaluation? }

Hereafter, we illustrate how we intend to proceed to answer the three research questions and we summarize the results.

Recently, the authors took part in a study commissioned by GDI, to assess the risk of disinformation exposed by a set of Italian online media, representative of the online media national market by geographical distribution, circulation and political ideology. 
The methodology, an overview of the contextual scenario and the results of the study are available at \url{https://bit.ly/3L9tBmi}. 

In addition, the authors own a valid NewsGuard license, for which they have access to the so-called `Nutrition Labels' of the news media, outcome of the NewsGuard's evaluation procedure.

By virtue of the above, we can compare two media rankings: the one obtained via our study for GDI, and the one obtained using the NewsGuard's labels\footnote{Because of contractual agreements, we cannot report in plain text the ranking obtained according to the GDI methodology, nor any data that could make it inferable. We can, however, conduct the comparison by reporting the differences found in the two rankings. In addition, the work that led to the actual results was conducted under the terms of the NewsGuard license.}.

Of the 31 online media analyzed, 7 have a different GDI score than the one obtained according to Newsguard. In particular, all 7 media result reliable for NG and not reliable for GDI.
To understand why scores were so different, we went down to the level of the evaluation of the individual criteria. To do this analysis, we initially mapped  GDI criteria and NG criteria, which we believe may also be useful for further investigation by Academia. 

After the conceptual mapping, we looked at which criteria the two organizations rated differently.  These are criteria that concern the policies that the media puts in place to ensure its transparency (e.g., the existence of declarations of editorial independence, declarations on funding sources and ownership structure). 
So, on the one hand, we can conclude that the concordance on criteria that mainly concern the analysis of news (discordance between title and text, use of sensationalist language, specific words or punctuation in the text...) lays a solid foundation to attempt the automated approach. In fact, we can think of using state-of-the-art NLP tools for the evaluation of such criteria. 
On the other hand, the discordance on criteria regarding presence of editorial line statements or ownership structure seems very odd, and probably the reason for this dissonance must be sought in the annotation campaign that each of the two agencies carries on. This opens the door for further investigations on the objectivity/subjectivity of annotation and crowd sourcing campaigns, which are currently discussed in works such as \cite{DBLP:journals/ipm/SopranoRBCSMD21,DBLP:journals/corr/abs-2107-11755}.

\paragraph{Contributions} This work brings the following contributions:

\begin{itemize}
\item The realization of a conceptual mapping between the criteria used by two agencies that have the same goal: to evaluate the degree of transparency and credibility of an online newspaper;
\item An analysis of the level of agreement in the evaluation of the same set of publications by different agencies; 
\item 
The basis for starting to think about which criteria can be calculated automatically, so as to partially automate the evaluation process and embrace a much larger number of news media. 
\end{itemize}

Overall, we believe that our study contributes to new directions in the automation of reliability verification, expanding the object of verification automation from the individual post/news to the news source.

\section{Useful Notions}
Here, we remind the reader the criteria used by the two to evaluate the news outlets, as well as their scoring systems.

\subsection{GDI and NG Criteria}

Tables~\ref{tab:GDI_criteria} and~\ref{tab:NG_criteria} (see the Appendix) show the list of criteria considered by the two organizations for evaluating a news site. In particular, Table~\ref{tab:GDI_criteria} reports the list of the GDI criteria, as they appear on the GDI website and in many of the GDI reports. 

The first column is the criterion name, the second one reports an abbreviation for that name (coined by us), the last column defines the criterion. Sometimes, a criterion is split into more subcriteria, reported in the third column. 

Table~\ref{tab:NG_criteria} shows the same information for the NG criteria. Even in this case, the criteria are public available on the NG website\footnote{\url{https://www.newsguardtech.com/ratings/rating-process-criteria/}}. The differences here are that criteria are not split in subcriteria, and the last column reports the score the news media gets when it meets the criterion. 

Both organizations consider 2 categories of criteria, one category that is more about the content and presentation of the individual news story, and one category that is about the editorial procedures and policies of the online media. 

Regarding GDI, in Table~\ref{tab:GDI_criteria} the first 9 criteria are related to content (e.g., the assessment of how well a headline reflects the content of the article, the presence or absence of a fact-based lede, the use of sensationalist terms in the article). The other 6 criteria relate to the media as a whole (e.g., the presence of statements of editorial independence, the attestation of the newspaper's editorial and financial structure, the presence of statements on sources of funding). 

With regard to NG, the first 5 criteria in Table~\ref{tab:NG_criteria} concerns the analysis of the content of the single article, while the remaining 4 criteria concern the transparency of the whole newspaper.


\subsection{GDI and NG scoring systems}\label{sec:scoringsystems}


The evaluation of the criteria leads both organizations to produce a final score for each news outlet\footnote{Computation of the final score, for both agencies, is described on their websites: \url{https://disinformationindex.org/wp-content/uploads/2019/12/GDI_Index-Methodology_Report_Dec2019.pdf}; \url{https://www.newsguardtech.com/ratings/rating-process-criteria/}.}. The score expresses how well the news source meets their criteria and, thus, fulfills good editorial principles and practices. Then, pre-defined threshold values  determine the quality  of the source under investigation: 
\begin{itemize}
    \item GDI category risk: Each media obtain an overall score as a result of the criteria evaluation. The news sources are then classified on the basis of a five-category risk scale based on the overall score. The risk categories were defined based on the distribution of risk ratings from 180 sites across six media markets in September 2020. This cross-country dataset was standardised to fit a normal distribution with a mean of 0 and a standard deviation of 1. The standardised scores and their distance from the mean were used to determine the bands for each risk level (i.e., minimum, low, medium, high, maximum). These bands are then used to categorise the risk levels for sites in each subsequent media market analysis. On a maximum score of 100, online media with a score $<$40 are labeled as `high risk of exposure to disinformation', between 40 and 50 a medium-high risk, between 50 and 60 a medium risk, between 60 and 70 a medium-low risk, and those with a score $>$70 pass the assessment positively. 
    \item NG overall score: NG uses 9 criteria to evaluate a news source. The total score the source can obtain is 100. Each criterion is associated with a numerical value, and the value assignment is all or nothing (criterion satisfied = associated value; criterion not satisfied = 0). Obviously, the sum of the values assigned to the 9 criteria is 100. 
    A news site with a score of 60 points or higher receives a positive rating in terms of credibility and transparency. A site with a score lower than 60 points receives a negative rating.
    A news site that achieves a score greater than 60 can still fail in one or more of the 9 criteria. The Nutrition Label provided by NG details which criteria are met and which are not.
\end{itemize}

\section{Methodology}
The methodology of the analysis will lead to two types of comparison regarding the procedures implemented by the two agencies for the evaluation of the same set of news outlets. 

The first comparison will concern the results of the evaluation, that is, we will measure the amount of news sources judged (i) the same way (i.e., judged reputable by both GDI and NG), (ii) in disagreement (i.e., reputable by GDI but not reputable by NG, and vice-versa). To do this we will consider (i) the overall scores of NG and GDI (See Section `GDI and NG scoring systems') and  (ii) the thresholds applied by both NG and GDI to the overall scores for identifying not reliable sources. The first comparison will give an answer to research question \textbf{RQ1}. 

The second comparison will address the criteria:
we will evaluate the amount of GDI criteria that find conceptual mapping with NG criteria, and vice versa. To do this, we will create an association matrix (GDI criteria - NG criteria), based on the specifications provided by the two organizations. This will allow us to answer question \textbf{RQ2}. 

Finally, for the criteria that find mapping, we will measure the level of agreement, i.e., that is, how much each GDI criterion has been evaluated in the same way as its homonym NG criterion (and vice-versa). This will answer question \textbf{RQ3}. 

\paragraph{Dataset}
The list of online Italian media considered in this article is a subset of the 34 online news outlets evaluated in the report  `Disinformation risk assessment: The online news market in Italy'~\cite{mediamarketItaly2022}.
The 34 news outlets have been chosen as representative of the Italian online publishing landscape. In particular, GDI 
proposed a first selection, based on the sites' reach (Alexa rankings, Facebook followers, and Twitter). Then, the set was thinned to arrive to a balanced group in terms of diffusion (either national or local), geographical location (i.e., North, Centre, or South and islands Italy) and political orientation (ultra-right, right, mainly neutral, left, ultra-left). 
Finally, for this work, we focus on 31 of the 34 outlets, because they were tagged both by NewsGuard and by GDI.

\section{Evaluation results}
\begin{table}[h!]
\begin{center}
\begin{tabular}{lcc}
& GDI Neg & GDI Pos \\
\hline
NG Neg   & 4 & 0 \\
NG Pos & 7 & 20 \\ 
\hline
\end{tabular}
\end{center}
\caption{Agreement and discordance in the investigated media ratings. One threshold value (50 for GDI, 60 for NG).}
\label{tab:rq1_gdi_high}
\end{table}

To answer Research Question \textbf{RQ1}, we consider the scores obtained by the 31 online media according to the GDI and NewsGuard assessments. 
It is important to note that NewsGuard applies a sharp division between trustworthy and untrustworthy online media (out of a maximum score of 100 points, those scoring $\ge$60 pass the assessment positively). In contrast, GDI divides media into 5 bands of risk of exposure to disinformation.

In order to compare the two rankings, we simply considered a two-class division for both the organizations. We proceeded as follows: we mapped the 5 GDI risk levels
%
%
on these 2 NG classes. In particular, the NG untrustworthy class was associated with the  maximum and high disinformation risk levels; the NG trustworthy class was associated with the medium, low and minimum risk levels. 
While the two-class division makes the evaluation less granular, it also makes the two rankings comparable, and we maintain the same thresholds originally chosen by the two associations,  
i.e., 60 for NG and 50 for GDI (50 is the threshold established by GDI to divide the high and medium risk levels).

Table~\ref{tab:rq1_gdi_high}  reports the number of sites that received concordant/discordant ratings from both organizations. Out of 31 sites, 20 received a positive rating from both, 7 received a negative rating from both, and 4 received a discordant rating. Notably, the latter were rated positively by NG, and negatively by GDI. None of the 31 sites received a positive rating from GDI and a negative rating from NG. As a percentage, 77.4\% of ratings are in agreement.

\subsection{Mapping of criteria}
We proceed with the analysis by looking for the two agencies' criteria there were evaluated in a convergent/divergent way. To reach the scope, we construct a conceptual mapping between criteria. This task has been carried out by the two of us, aided by our knowledge of the GDI criteria, which we evaluated on each of the 31 news outlets, see~\cite{mediamarketItaly2022}.
We initially conducted the task on our own, associating GDI criteria with NG criteria, and vice-versa, based on their textual description and our experience.  It took us about three hours to finish the task. 
Next, we compared our associations, dwelling on those for which we disagreed.
%
For the few discordant associations, we  managed to find  a common view and, thus, an agreement. This second phase lasted about two hours.


Table~\ref{tab:mapping} shows the result. Labels $S$ and $W$ stand for Strong and Weak. We assigned a Strong connection between criteria when the definitions were such that they were virtually the same. This is the case, for example, of GDI: {\it Headline Accuracy} and NG: {\it Avoid Deceptive Headlines}. 

A Weak label was instead assigned when only part of the definition of a criterion by one agency is matched by the definition of another criterion defined by the other agency. For example, GDI: {\it Common Coverage} is, according to its definition, `indicative of a true and significant event' and, thus, it finds a connection with NG: {\it Does not repeatedly publish false content}, even if in a less direct way.

As shown in Table~\ref{tab:GDI_criteria}, some of the GDI criteria find declination in multiple subcriteria. An asterisk in the mapping indicates that a NG criterion has a partial match with some of the subcriteria of a GDI criterion.

Moreover, to give an idea about the strength of the conceptual link found between each criterion of organization $x$ mapped onto one (or more) criteria of organization $y$, we introduce the concept of {\it Conceptual Mapping Level} (CML). In particular, we have defined the following four CML levels of mapping, based on the proportion of Strong and Weak mappings identified for each criterion:
\begin{itemize}
    \item 4, Strong: There is perfect conceptual mapping between 1 criterion of organization $x$ and 1 (or more) criteria of organization $y$;
    \item 3, Almost Strong: If a criterion of organization $x$ finds mapping in more than one criterion of organization $y$, at least half of these mappings are Strong;
    \item 2, Almost Weak: If a criterion of organization $x$ finds mapping in more than one criterion of organization $y$, and more than half of these mappings are Weak;
    \item 1, Weak: There is no perfect conceptual mapping between 1 criterion of organization $x$ and 1 (or more) criteria of organization $y$.
    \end{itemize}

\newcommand*\rot{\rotatebox{90}}

\begin{table*}[h!]
\begin{center}
\begin{tabular}{c|ccccccccc|c}
 & \rot{RepFalseCont} & \rot{InfoResp} & \rot{ErrCorr} & \rot{NewsOpDiff} & \rot{AvDecHeadlines} & \rot{DiscOwnFin} & \rot{LabAds} & \rot{RevConfOfInt} & \rot{ContCreators} & \rot{GDI to NG CML} \\
 \hline

ArtBias &  & $S$ &  & $S$ &  &  &  &  &  & $4$ \\
Bylnfo &  & $W$ &  &  &  &  &  &  & $S$ & $3$ \\
ComCov & $W$ &  &  &  &  &  &  &  & & $1$ \\
HeadAcc &  &  &  &  & $S$ &  &  &  & & $4$ \\
LedePres &  & $S$ &  &  &  &  &  &  & & $4$ \\
NegTarg &  & $S$ &  &  &  &  &  &  & & $4$ \\
RecCov & $W$ & $W$ &  &  &  &  &  &  & & $1$ \\
SensLang &  & $S$ &  &  & $W$ &  &  &  & & $3$ \\
VisPres &  & $S$ &  &  &  &  &  &  & & $4$ \\
Attr &  & $S$ &  &  &  &  &  &  & $W$ & $3$ \\
CommPol &  &  &  &  &  &  &  &  & &  - \\
EdPrincPract &  & $S^{*}$ &  & $S^{*}$ &  & $S^{*}$ &  & $S^{*}$ & & $4^{*}$ \\
EnsAcc & $W$ & $S^{*}$ & $S^{*}$ &  &  &  &  &  & & $3^{*}$ \\
Fund &  &  &  &  &  & $S$ & $W^{*}$  & $W^{*}$ & & $2^{*}$ \\
Own &  &  &  &  &  & $S$ &  & $S^{*}$ & & $4^{*}$ \\ \hline
NG to GDI CML &  $1$ & $3^{*}$ & $4^{*}$ & $4^{*}$ & $3$ & $4^{*}$ & $1^{*}$ & $3^{*}$ & $3$ &  \\ 
\end{tabular}
\end{center}
\caption{Conceptual mapping between GDI and NG criteria (in the left column and header, respectively). The degree of mapping can be `strong' (S) or `weak' (W). When a GDI criterion is composed of multiple subcriteria, the S or W assignment is by majority vote. The asterisk indicates that a NG criterion has been partially mapped into a GDI one (i.e., we found a partial conceptual mapping with only some of the GDI subcriteria from which it is composed.). Side numbers summarize the {\it Conceptual Mapping Level} (CML), explained in the 'Mapping Criteria section'.}
\label{tab:mapping}
\end{table*}

The results of the conceptual mapping (see Table~\ref{tab:mapping}) can be summarised as follows:

\begin{itemize}
    \item All NG criteria find at least one mapping to GDI criteria. Seven ({\it InfoResp, ErrCorr, NewsOpDiff, AvDecHeadlines, DiscOwnFin, RevConfOfInt, ContCreators}) reach at least the `Almost Strong' CML and three of these ({\it ErrCorr, NewsOpDiff, DiscOwnFin}) reach the `Strong' one. Six criteria have at least a partial mapping; three of these ({\it ErrCorr, NewsOpDiff, RevConfOfInt}) with at least half of partial mappings. Only two criteria ({\it RepFalseCont, LabAds}) reach the `Weak' value of CML.
    \item Fourteen GDI criteria (out of fifteen ones) find at least one mapping to NG criteria. The only one for which we cannot find a mapping is the criterion regarding the presence, on the website, of policies regulating the user-generated content: the {\it Comments policies}. Eleven criteria ({\it ArtBias, ByInfo, HeadAcc, LedePres, NegTarg, SensLang, VisPres, Attr, EdPrincPract, EnsAcc, Own}) reach at least the `Almost Strong' value of CML. Four criteria ({\it EdPrincPract, EnsAcc, Fund, Own}) have at least half of partial mappings. Only three criteria ({\it ComCov, RecCov, Fund}) reach a value of CML less (or equal) than `Almost Weak'.
\end{itemize}
 
Recalling the question \textbf{RQ2}, where we wondered whether the criteria of the two organizations could be mapped onto each other, the answer is positive. Some mappings are stronger than others in the sense that we find a precise match in the definition of the criteria. Still, in the end, (i) all the criteria are matched (apart from the GDI criterion on moderation of user-generated content) and (ii) the majority of the criteria (for both GDI and NG) achieves a CML value of at least `Almost Strong'. These positive signals show that the two agencies are moving along similar guidelines for assessing the credibility and transparency of an online media outlet. 
In the next section, we will seek to answer the third research question, regarding the evaluation of the criteria agreement.

\subsection{Evaluation of the agreement on the single criteria}

Regardless of how a source was classified (low/high reliable by either GDI or NG or both), here we analyse how much agreement there is between the criteria that find a conceptual mapping,   considering all the media sources in the study. 
 
\begin{figure}[ht!]
  \includegraphics[width=\linewidth]{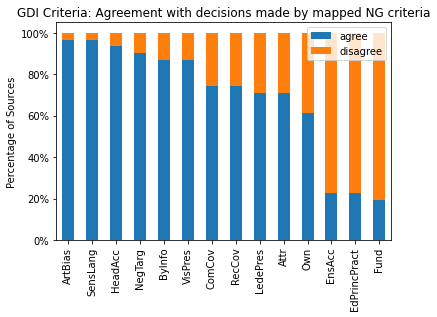}
  \caption{For each GDI criterion, percentage of online media rated the same under the analogous NG criteria.}
  \label{fig:ng_total_agreement}
\end{figure}

Thus, for each GDI criterion and the NG criteria that find conceptual mapping to it, Figure~\ref{fig:ng_total_agreement} shows the level of agreement between the evaluation of the GDI criterion and that of its NG `analog' ones.

On the x-axis, we list each GDI criterion. The y-axis shows the percentage of the investigated online media rated the same under the analogous NG criteria.  For example, the NG analogs of the {\it Article Bias} criterion, which aims to assess whether a news story is written in neutral or biased terms, were evaluated the same way on 30 media out of 31 ones (in percentage terms, is 96.77\%).

The GDI criteria whose NG analogues were evaluated more discordantly concern the presence of statements about the ownership structure of the media ({\it Own}); the implementation of pre-publication fact-checking and post-publication error correction procedures ({\it EnsAcc}); the presence of independence declarations on the media website ({\it EdPrincPract}) and the declarations about funding sources ({\it Fund}). For the last 3 GDI criteria, ~20\% of the analyzed media only (i.e., 6 media) feature the same evaluation of the analogous NG criteria. 

\begin{figure}
  \includegraphics[width=\linewidth]{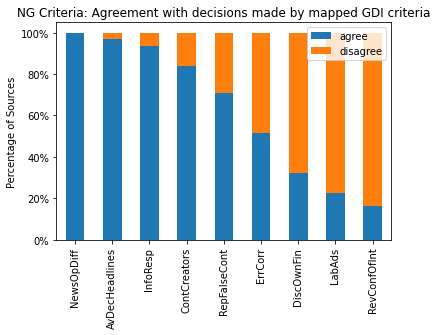}
  \caption{For each NG criterion, percentage of online media rated the same under the analogous GDI criteria.}
  \label{fig:gdi_total_agreement}
\end{figure}

Figure~\ref{fig:gdi_total_agreement} shows the same analysis, starting from each NG criterion. 

What we notice is that the biggest disagreement in the evaluation of similar criteria is always on those that concern the investigation of the whole newspaper, and not on those that concern the evaluation of the single news item. 
In fact, even in the case of Figure~\ref{fig:gdi_total_agreement}, the similar GDIs of ownership, funding sources, editorial structure and guidelines (\textit{DiscOwnFin, ContCreators, RevConfOfInt}) were evaluated in the same way in well under 40\% of the considered media. 

\begin{table}[!ht]
    \centering
    \begin{tabular}{lccc}
        \bf{Criteria} & \bf{Cat.} & \bf{CML} & \bf{Agreement(\%)}  \\ 
        \midrule
        \multicolumn{4}{l}{\bf{GDI}}\\
        \rowcolor{lightgray}
        ArtBias & CR & 4 & 96.77  \\ 
        \rowcolor{lightgray}
        SensLang & CR & 3 & 96.77  \\ 
        \rowcolor{lightgray}
        HeadAcc & CR & 4 & 93.55  \\ 
        \rowcolor{lightgray}
        NegTarg & CR & 4 & 90.32  \\ 
        \rowcolor{lightgray}
        Bylnfo & CR & 3 & 87.1  \\ 
        \rowcolor{lightgray}
        VisPres & CR & 4 & 87.1  \\ 
        ComCov & CR & 1 & 74.19  \\ 
        RecCov & CR & 1 & 74.19  \\ 
        \rowcolor{lightgray}
        LedePres & CR & 4 & 70.97  \\ 
        \rowcolor{lightgray}
        Attr & T & 3 & 70.97  \\ 
        Own & T & 4* & 61.29  \\ 
        EnsAcc & T & 3* & 22.58  \\ 
        EdPrincPract & T & 4* & 22.58  \\ 
        Fund & T & 2* & 19.35  \\ 
        \midrule
        \multicolumn{4}{l}{\bf{NG}}\\
        NewsOpDiff & CR & 4* & 100  \\ 
        AvDecHeadlines & CR & 3 & 96.77  \\ 
        InfoResp & CR & 3* & 93.55  \\ 
        ContCreators & T & 3 & 83.87  \\ 
        RepFalseCont & CR & 1 & 70.97  \\ 
        ErrCorr & CR & 4* & 51.61  \\ 
        DiscOwnFin & T & 4* & 32.26  \\ 
        LabAds & T & 1* & 22.58  \\ 
        RevConfOfInt & T & 3* & 16.13 \\ 
        \midrule
    \end{tabular}
    \caption{Relation between the {\it Conceptual Mapping Level} (CML) and the agreement on the criteria evaluation. Category=\{CR=credibility, T=transparency, \}; The asterisk on the CML value indicates at least one partial mapping.}
    \label{tab:agreement}
\end{table}

To give more insights about the agreement values, we also study the relation between the agreement  and the level of mapping (see Section `Mapping of criteria'). Table~\ref{tab:agreement} lists the GDI and NG criteria, the category to which they belong (T=transparency of media/CR=credibility of published news), the level obtained in the conceptual mapping (see also CML values in Table~\ref{tab:mapping}), and finally, the degree of agreement with the mapped criteria of the other organization (same values as in Figure~\ref{fig:ng_total_agreement} and Figure~\ref{fig:gdi_total_agreement}). 

We observe that the GDI criteria that obtain a highest agreement (i) belong to the category {\it credibility}, and (ii) have no partial conceptual mapping (no asterisk). We recall that partial mapping (presence of asterisk) indicates that the criterion  finds mapping but is only partially represented by the other agency. It is, therefore, reasonable to think that, in presence of partial mapping, the agreement in the criteria evaluation is lower, since some specific aspects of one criterion are not taken into account by the set of criteria on which it is mapped. 

NG criteria with the highest CML values belong that to the {\it credibility} category.
Most of the criteria with high agreement find at least a partial mapping:  this is true for {\it NewsOpDiff} and {\it InfoResp} and means that, although the mapping is partial, these two criteria are still well represented by the mapped criteria. This assumption is also supported by the high CML values achieved by {\it NewsOpDiff} and {\it InfoResp} (Strong and Almost Strong,  respectively) and a large number of mapped criteria for {\it InfoResp}.

\section{Discussion}\label{sec:discussion}
This work was born from the question of whether and to which extent it is possible to automate the current journalistic procedures of evaluation of an online media, based on criteria that consider aspects of credibility of the published news and transparency of the media itself. 

Moving towards automated classification of news media --a {\it kind of} automated fact checking, but on the whole news outlet rather than the single piece of information-- 
is useful since, to date, many online media have not yet been manually evaluated. This is certainly true for the long tail of news outlets characterized by low traffic. However, even in considering the 34 Italian media representative of the country's information landscape, 3 of them do not have a NewsGuard rating, at time of writing.  

However, in our opinion, it is useless to start with the automatic computation of certain features, if we do not first carry out an assessment of the solidity of the manual procedures implemented to date. 
Hence, we started by assessing how many, out of a fixed set of media, were considered the same way by two well-known organizations of journalists and media experts working to unveil low reliable news sources. 

It was possible to have the ranking of the analysed online media  because part of the authors conducted the study on behalf of GDI.
For NewsGuard,  we hold a NG license at time of writing.

The first analysis carried out was to quantify the number of sites judged in agreement by both organizations. A coarse-grained analysis (i.e., using only 1 threshold to distinguish `good' news media from `bad' news media) found that 20 sites were judged negatively by both the agencies, 7 positively, while for 4 sites they gave discordant judgments. This addresses research question \textbf{RQ1}. 

The second analysis led to the definition of a conceptual mapping between the GDI and NG criteria. Mostly based on the literal definitions of the criteria, we found that the criteria have a correspondence, in some cases strong (the definitions of the criteria are essentially the same), in some cases weak (the definition of a GDI criterion includes or is included in the definition of an NG criterion). In addition, the correspondence may be partial when the definition of a NG criterion includes, or is included in, some of the sub-criteria into which a GDI criterion is divided. The construction of the mapping answers research question \textbf{RQ2}. 

Finally, we asked which of all the mapped criteria were evaluated the same by the two organizations, on the largest (or smallest) number of online media. The outcome of the analysis was, in our opinion, noteworthy. The criteria regarding the evaluation of the single news item (such as, e.g., the appropriateness of the title, the presence of the author's name, the presence of a fact-based lede, the type of language, the visual presentation, etc.) are the criteria whose evaluation is agreed upon on all, or many, of the analyzed news outlets. 
There is a greater discordance in the evaluation of criteria concerning  the news media as a whole. In particular, one agency, GDI, rated some media more negatively on these criteria than the other. 
This answers research question \textbf{RQ3}, and brings food for thought. 

On the one hand, the discrepancy in the rating may be due to how annotators were instructed by the organizations employees. On the other hand, there is an underlying bias on annotation processes. Considering different groups of annotators (or samples of articles) may lead to different annotations and, consequently a possible different assessment of the reliability for the same news source.
Work in~\cite{bhuiyan2020investigating} highlights that
  among groups of experienced annotators with different backgrounds (i.e., journalists and scientists), there is no perfect agreement regarding the credibility assessment of news. 
%
Therefore, it is reasonable to think that in case organizations {\it a la} GDI and NG commission the analysis to people with different skills, the result of the evaluation changes even if the criteria find a perfect conceptual mapping.
A promising direction to improve the annotation process is to set up crowd-sourcing campaigns formed by persons with different expertise and then compare the result of the evaluations with the judgment of experienced journalists. Once evaluated which of the campaigns gives a result more similar to that given by the experts, the adoption of ad-hoc campaigns could assist the work of these journalistic organizations, given their difficulty in scaling up.
Furthermore, crowd-sourcing based projects have recently been exploited also by technology companies such as Meta and Twitter to review and rate viral misinformation (see, e.g., \cite{fcFacebook}, \cite{fcTwitter}). Thus, the move to use such campaigns to judge the source of news appears promising.

\paragraph{Summing up:} The meta-question we asked in the introduction of this manuscript was: {\it Is it worth trying to automate the evaluation process carried out manually so far?}
The analyses help answer that. First of all, we have seen how it is possible to construct a conceptual mapping between criteria adopted by two different organizations, also highlighting the degree of correspondence between criteria (i.e., strong, weak, partial). Second, we analyzed which criteria were evaluated the same way by the organizations, on the same set of news media.

Both the existence of the mapping between criteria and the agreement on their evaluations aid the choice of which criteria/features might be worth automating.
We believe that a good starting point is to consider those criteria for which i) there is strong mapping (i.e., CML 4 and 3) and ii) the evaluations are more in agreement (e.g., the agreement is at least on 2/3 of the media that, reported in percentage, is equal to approximately 67\%.

In Table~\ref{tab:agreement} we have highlighted in grey the GDI criteria which, in relation to the thresholds we have set (CML $\ge$ 3, agreement percentage $\ge$  67), appear  as good candidates to be automatically computed. Obviously, the thresholds were set by us in a rational but arbitrary manner, and different thresholds may be considered in the future. 


Criteria matching and agreement in their evaluation has also another implication. Suppose that, using state-of-the-art NLP tools, we automatically compute the values of some GDI criteria such as, for example, the use of sensational language in the news and the accuracy of the headline (SensLang and HeadAcc in Table~\ref{tab:agreement}). The fact of having correspondence and agreement in the evaluation with the NG criterion `Avoid Deceptive Headlines' (see both the mapping in Table~\ref{tab:mapping} and the agreement figures --Figure~\ref{fig:gdi_total_agreement} and \ref{fig:ng_total_agreement}) means that the value of `Avoid Deceptive Headlines' can be taken as a reference, even as ground truth, to evaluate the goodness of the result of the automatic evaluation of the corresponding GDI criteria.

Finally, we would like to point out that our choice to start with the automatic computation of GDI criteria is due to the fact that in our previous work on the assessment of the media market in Italy we have gained a good understanding of the meaning of these criteria and how to evaluate them.

\paragraph{Limitations:}
In this manuscript, we have explored the -mostly manual- procedures performed by organizations experienced in journalism and media communication to assess the degree of transparency and credibility of a news source. 
The analysis was carried out with the ultimate aim of 1) emerging a set of criteria on which to focus a first automatic computation process; 2) better investigating the reasons why on some criteria there is no agreement.
Obviously, this work is not without limitations. 
First of all, the analysed procedures are only two, as well as the news media are Italian and small in number. 
Given the difference in classification of the two agencies (NG has two rating levels, GDI has 5), we opted for a binary classification in the comparison, losing granularity. However, setting arbitrary thresholds to split the NewsGuard ranking into several parts seemed to us too arbitrary. 
In constructing the conceptual mapping between criteria, it was not always possible to establish a strong connection. In addition, the lack of agreement we found may be due to: 1) a different process of implementation of the criterion, dictated by the organizations themselves; 2) an inherent bias in the annotators (e.g., the study on the media market in Italy was conducted by computer scientists, the study on the same media by NewsGuard may have been conducted by a group of annotators with a different background, e.g., journalists); 3) the articles chosen to assess the credibility aspects were obviously not the same in the two studies; 4) the study of the news website has been realized at different times (its content may have changed and the two assessments may have been affected by this change). 

Aware of these limitations, we believe that it is noteworthy that the two procedures lead to the same outcomes, at least in terms of news content. Furthermore, it is interesting that criteria such as, e.g., the existence of declarations of independence, or declarations on ownership and editorial structure of the media are the criteria for which the two procedures gave opposite results. A good line of inquiry for the future is to study how different annotation campaigns lead to the same results for news source evaluation. We argue that the outcome of that analysis will improve current assessment procedures. 

\section{Conclusions}
Driven by the curiosity to understand if and to what extent a process of evaluating a news source can be made automatic, in this paper we compared 2 journalistic procedures currently in use to classify new sites as reliable or not. For the aspect concerning the published content, we found a good correspondence in the evaluations of the criteria taken into consideration by the two procedures. The situation is different for the criteria regarding the transparency of the source as a whole. This opens the way to new investigations, such as, for example, on the possibility of calibrating ad hoc crowd-sourcing campaigns to better evaluate those aspects that find discordant evaluation.

\section{Acknowledgments}
The authors acknowledge support from the Project TOFFEe (TOols for Fighting FakEs), funded by IMT School for Advanced Studies Lucca.


\begin{thebibliography}{}

\bibitem[\protect\citeauthoryear{Aker, Vincentius, and
  Bontcheva}{2019}]{aker2019credibility}
Aker, A.; Vincentius, K.; and Bontcheva, K.
\newblock 2019.
\newblock Credibility and transparency of news sources: Data collection and
  feature analysis.
\newblock In {\em NewsIR@ SIGIR},  15--20.

\bibitem[\protect\citeauthoryear{Bhuiyan \bgroup et al\mbox.\egroup
  }{2020}]{bhuiyan2020investigating}
Bhuiyan, M.~M.; Zhang, A.~X.; Sehat, C.~M.; and Mitra, T.
\newblock 2020.
\newblock Investigating differences in crowdsourced news credibility
  assessment: Raters, tasks, and expert criteria.
\newblock {\em Proceedings of the ACM on Human-Computer Interaction}
  4(CSCW2):1--26.

\bibitem[\protect\citeauthoryear{Caldarelli \bgroup et al\mbox.\egroup
  }{2021}]{CaldarelliNPPS21}
Caldarelli, G.; {De Nicola}, R.; Petrocchi, M.; Pratelli, M.; and Saracco, F.
\newblock 2021.
\newblock Flow of online misinformation during the peak of the {COVID-19}
  pandemic in italy.
\newblock {\em {EPJ} Data Sci.} 10(1):34.

\bibitem[\protect\citeauthoryear{Grinberg \bgroup et al\mbox.\egroup
  }{2019}]{doi:10.1126/science.aau2706}
Grinberg, N.; Joseph, K.; Friedland, L.; Swire-Thompson, B.; and Lazer, D.
\newblock 2019.
\newblock Fake news on {T}witter during the 2016 {U.S.} presidential election.
\newblock {\em Science} 363(6425):374--378.

\bibitem[\protect\citeauthoryear{Mattei \bgroup et al\mbox.\egroup
  }{2022}]{Mattei22bowties}
Mattei, M.; Pratelli, M.; Caldarelli, G.; Petrocchi, M.; and Saracco, F.
\newblock 2022.
\newblock Bow-tie structures of {Twitter} discursive communities.
\newblock {\em CoRR} abs/2202.03316.

\bibitem[\protect\citeauthoryear{Meta}{2022}]{fcFacebook}
Meta.
\newblock 2022.
\newblock Meta journalism project.
\newblock [Online; accessed 08-April-2022].

\bibitem[\protect\citeauthoryear{Petrocchi, Pratelli, and
  Spognardi}{2022}]{mediamarketItaly2022}
Petrocchi, M.; Pratelli, M.; and Spognardi, A.
\newblock 2022.
\newblock Disinformation risk assessment: The online news market in italy.
\newblock Online: https://bit.ly/3L9tBmi.

\bibitem[\protect\citeauthoryear{Roitero \bgroup et al\mbox.\egroup
  }{2021}]{DBLP:journals/corr/abs-2107-11755}
Roitero, K.; Soprano, M.; Portelli, B.; Luise, M.~D.; Spina, D.; Mea, V.~D.;
  Serra, G.; Mizzaro, S.; and Demartini, G.
\newblock 2021.
\newblock Can the crowd judge truthfulness? {A} longitudinal study on recent
  misinformation about {COVID-19}.
\newblock {\em CoRR} abs/2107.11755.

\bibitem[\protect\citeauthoryear{Shao \bgroup et al\mbox.\egroup
  }{2018}]{Shao18}
Shao, C.; Ciampaglia, G.; Varol, O.; Yang, K.-C.; Flammini, A.; and Menczer, F.
\newblock 2018.
\newblock The spread of low-credibility content by social bots.
\newblock {\em Nature Communications} 9.

\bibitem[\protect\citeauthoryear{Soprano \bgroup et al\mbox.\egroup
  }{2021}]{DBLP:journals/ipm/SopranoRBCSMD21}
Soprano, M.; Roitero, K.; Barbera, D.~L.; Ceolin, D.; Spina, D.; Mizzaro, S.;
  and Demartini, G.
\newblock 2021.
\newblock The many dimensions of truthfulness: Crowdsourcing misinformation
  assessments on a multidimensional scale.
\newblock {\em Inf. Process. Manag.} 58(6):102710.

\bibitem[\protect\citeauthoryear{Twitter}{2022}]{fcTwitter}
Twitter.
\newblock 2022.
\newblock Empowering people on {T}witter to create a better-informed world.
\newblock [Online; accessed 08-April-2022].

\bibitem[\protect\citeauthoryear{Zeng, Abumansour, and
  Zubiaga}{2021}]{DBLP:journals/llc/ZengAZ21}
Zeng, X.; Abumansour, A.~S.; and Zubiaga, A.
\newblock 2021.
\newblock Automated fact-checking: {A} survey.
\newblock {\em Lang. Linguistics Compass} 15(10).

\bibitem[\protect\citeauthoryear{Zubiaga and
  Jiang}{2020}]{DBLP:journals/tweb/ZubiagaJ20}
Zubiaga, A., and Jiang, A.
\newblock 2020.
\newblock Early detection of social media hoaxes at scale.
\newblock {\em {ACM} Trans. Web} 14(4):18:1--18:23.

\end{thebibliography}
\newpage
\onecolumn

\section{Appendix}

\begin{table}[H]
\small
\centering
\begin{tabular}{ p{0.2\linewidth} c p{0.4\linewidth} c }
\textbf{Criteria name} & \textbf{Short name} & \textbf{Definition} & \textbf{points} \\ \hline
Does not repeatedly publish false content & RepFalseCont & The site does not repeatedly produce stories that have been found - either by journalists at NewsGuard or elsewhere - to be clearly and significantly false, and which have not been quickly and prominently corrected. & 22 \\ \hline
Gathers and presents information responsibly & InfoResp & Content providers are generally fair and accurate in reporting and presenting information. They reference multiple sources, preferably those that present direct, firsthand information on a subject or event or from credible second hand news sources, and they do not egregiously distort or misrepresent information to make an argument or report on a subject. & 18 \\ \hline
Regularly corrects or clarifies errors & ErrCorr & The site makes clear how to report an error or complaint, has effective practices for publishing clarifications and corrections, and notes corrections in a transparent way. & 12.5 \\ \hline
Handles the difference between news and opinion responsibly & NewsOpDiff & Content providers who convey the impression that they report news or a mix of news and opinion distinguish opinion from news reporting, and when reporting news, do not egregiously cherry pick facts or stories to advance opinions. Content providers who advance a particular point of view disclose that point of view. & 12.5 \\ \hline
Avoids deceptive headlines & AvDecHeadlines & The site generally does not publish headlines that include false information, significantly sensationalize, or otherwise do not reflect what is actually in the story & 10 \\ \hline
Website discloses ownership and financing & DiscOwnFin & The site discloses its ownership and/or financing, as well as any notable ideological or political positions held by those with a significant financial interest in the site, in a user-friendly manner. & 7.5 \\ \hline
Clearly labels advertising & LabAds & The site makes clear which content is paid for and which is not. & 7.5 \\ \hline
Reveals who’s in charge, including possible conflicts of interest & RevConfOfInt & Information about those in charge of the content is made accessible on the site & 5 \\ \hline
The site provides the names of content creators, along with either contact or biographical information & ContCreators & Information about those producing the content is made accessible on the site & 5 \\ \hline
\end{tabular}
\caption{NewsGuard criteria specifications.}
\label{tab:NG_criteria}
\end{table}

\begin{table}[]
\small
\begin{tabular}{ p{0.15\linewidth} c p{0.15\linewidth} p{0.5\linewidth} }
\textbf{Criteria Name} & \textbf{Short name} & \textbf{Sub-indicators} & \textbf{Definition} \\ \hline
Headline accuracy & HeadAcc & - & Rating for how accurately the story’s headline describes the content of the story. Indicative of clickbait \\ \hline
Byline information & Bylnfo & - & Rating for how much information is provided in the article’s byline. Attribution of stories creates accountability for their veracity \\ \hline
Lede present & LedePres & - & Rating for whether the article begins with a fact-based lede. Indicative of fact-based reporting and high journalistic standards \\  \hline
Common coverage & ComCov & - & Rating for whether the same event has been covered by at least one other reliable local media outlet. Indicative of a true and significant event \\  \hline
Recent coverage & RecCov & - & Rating for whether the story covers a news event or development that occurred within 30 days prior to the article’s publication date. Indicative of a newsworthy event, rather than one which has been taken out of context \\  \hline
Negative targeting & NegTarg & - & Rating for whether the story negatively targets a specific individual or group. Indicative of hate speech, bias or an adversarial narrative. \\  \hline
Article bias & ArtBias & - & Rating for the degree of bias in the article. Indicative of neutral fact-based reporting or well-rounded analysis. \\  \hline
Sensational language & SensLang & - & Rating for the degree of sensationalism in the article. Indicative of neutral fact-based reporting or well-rounded analysis. \\  \hline
Visual presentation & VisPres & - & Rating for the degree of sensationalism in the visual presentation of the article. Indicative of neutral fact-based reporting or well-rounded analysis. \\  \hline
Attribution & Attr & - & Rating for the number of policies and practices identified on the site. Assesses policies regarding the attribution of stories, facts, and media (either publicly or anonymously); indicative of policies that ensure accurate facts, authentic media and accountability for stories. \\  \hline
\multirow{2}{*}{Comment policies} & \multirow{2}{*}{CommPol} & Policies & Rating for the number of policies identified on the site. Assesses policies to reduce disinformation in user-generated content. \\
 & & Moderation & Rating for the mechanisms to enforce comment policies identified on the site. Assesses the mechanism to enforce policies to reduce disinformation in user-generated content \\ \hline
\multirow{4}{*}{\shortstack{Editorial principles\\ and practices}} & \multirow{4}{*}{EdPrincPract} & Editorial independence & Rating for the number of policies identified on the site. Assesses the degree of editorial independence and the policies in place to mitigate conflicts of interest \\
 & & Adherence to narrative & Rating for the degree to which the site is likely to adhere to an ideological affiliation, based on its published editorial positions. Indicative of politicised or ideological editorial decision-making \\  
 &  & Content guidelines & Rating for the number of policies identified on the site. Assesses the policies in place to ensure that factual information is reported without bias \\
& & News vs. analysis & Rating for the number of policies and practices identified on the site. Assesses the policies in place to ensure that readers can distinguish between news and opinion content \\ \hline
\multirow{2}{*}{Ensuring accuracy} & \multirow{2}{*}{EnsAcc} & Pre-publication fact checking & Rating for the number of policies and practices identified on the site. Assesses policies to ensure that only accurate information is reported \\
 & & Post-publication corrections & Rating for the number of policies and practices identified on the site. Assesses policies to ensure that needed corrections are adequately and transparently disseminated \\ \hline
\multirow{3}{*}{Funding} & \multirow{3}{*}{Fund} & Diversified incentive structure & Rating for the number of revenue sources identified on the site. Indicative of possible conflicts of interest stemming for over-reliance on one or few sources of revenue \\
 &  & Accountability to readership & Rating based on whether reader subscriptions or donations are identified as a revenue source. Indicative of accountability for high-quality information over content that drives ad revenue. \\
& & Transparent funding & Rating based on the degree of transparency the site provides regarding its sources of funding. Indicative of the transparency that is required to monitor the incentives and conflicts of interest that can arise from opaque revenue sources. \\ \hline
\multirow{2}{*}{Ownership} & \multirow{2}{*}{Own} & Owner-operator division & Rating based on the number of distinct executive- or board-level financial and editorial decision-makers listed on the site. Indicative of a separation between financial and editorial decision-making, to avoid conflicts of interest. \\
 & & Transparent ownership & Rating based on the degree of transparency the site provides regarding its ownership structure. Indicative of the transparency that is required to monitor the incentives and conflicts of interest that can arise from opaque ownership structures.\\ \hline
\end{tabular}
\caption{GDI criteria specifications.}
\label{tab:GDI_criteria}
\end{table}

\end{document}